\documentstyle{l-aa}
\def\tvi(#1,#2){\vrule height #1pt depth #2pt width 0pt}
\def\p{\partial}

\def\tx{\tilde\xi}
\def\bk{\bar k}
\def\bx{\bar\xi}

\def\tom{\tilde\omega}
\def\tod{\tilde\omega_{\Delta}}

\def\e{{\rm e}}
\def\d{{\rm d}}
\def\ie{i.e. }
\def\eg{e.g. }
\def\etal{et al. }
\def\VAc{V_{\rm A}^2}
\def\VA{V_{\rm A}}
\newlength{\largeur}
\newlength{\saut}
\def\marge#1{
\setlength{\largeur}{\columnwidth}
\addtolength{\largeur}{-#1}
\setlength{\saut}{0.5\largeur}\hspace*{\saut}}
\def\picture #1 by #2 (#3){
 \marge{#1} \vbox to #2{
  \hrule width #1 height 0pt depth 0pt
  \vfill
  \special{picture #3}}}
\def\scaledpicture #1 by #2 (#3 scaled #4){{
 \dimen0=#1 \dimen1=#2
 \divide\dimen0 by 1000 \multiply\dimen0 by #4
 \divide\dimen1 by 1000 \multiply\dimen1 by #4
 \picture \dimen0 by \dimen1 (#3 scaled #4)}}
\begin{document}

\thesaurus{12
(02.01.2;
 02.09.1;
 02.13.2;
 09.03.2;
 11.09.4;
 11.13.2)}

\title{The Parker-Shearing instability in azimuthaly magnetized discs}

\author{T.~Foglizzo
\inst{1,2}
\and
M.~Tagger
\inst{1}}

\offprints{foglizzo@mpa-garching.mpg.de}

\institute {CEA/DSM/DAPNIA, Service d'Astrophysique, CE-Saclay, 91191
Gif-sur-Yvette Cedex, France
\and
Max-Planck-Institut f\"ur Astrophysik, Karl-Schwarzschild-Str. 1,
Postfach 15 23,
85740 Garching, Germany}

\date{Received June 3; accepted August 18, 1994}

\maketitle
%\maintitlerunninghead{The Parker-Shearing instability...}
%\authorrunninghead{Foglizzo \& Tagger}

\begin{abstract}

We describe the effects of both magnetic buoyancy and differential rotation
on a disc of isothermal gas embedded in a purely azimuthal magnetic field,
in order to study the evolution and interplay of Parker and shearing
instabilities.\\
We perform a linear analysis of the evolution of perturbations in the
shearing sheet model. Both instabilities occur on the slow MHD branch of the
dispersion relation, and can affect the same waves. We put a stress on the
natural polarization properties of the slow MHD waves to get a better
understanding of the physics involved. The mechanism of the shearing
instability is described in details.\\
Differential rotation can transiently stabilize slow MHD waves with a vertical
wavelength longer than the scale height of the disc, against the Parker
instability. Waves with a vertical wavelength shorter
than the scale height of the disc are subject to both the Parker and the
transient shearing instabilities. They occur in different ranges of radial
wavenumbers, \ie at different times in the shearing evolution ; these ranges
can overlap or, on the contrary, be separated by a phase of wave-like
oscillations, depending on the strength of differential rotation.\\
These analytical results, obtained in a WKB approximation, are found to be in
excellent agreement with numerical solutions of the full set of linearized
equations. Our results can be applied to both galactic and accretion discs.

\keywords{accretion discs --
instabilities --
MHD --
cosmic rays --
Galaxies: ISM --
Galaxies: magnetic fields}

\end{abstract}

\section{Introduction}

The Parker instability (Parker 1966) relies on the natural tendency of the
horizontal component of the magnetic flux to escape towards the surface of
a vertically stratified atmosphere (Kruskal \& Schwarzschild 1954,
Tserkovnikov 1960, Newcomb 1961, Yu 1966). This effect of magnetic buoyancy
occurs wherever a magnetic field is embedded in a density gradient: its
classical manifestations are the sunspots on the surface of the sun (Parker
1955), the escape of the magnetic field lines towards
the halo of disc galaxies (Parker 1966, Lachi\`eze-Rey \etal 1980), or the
formation of the molecular clouds which appear as ``beads on a string" in
galaxies (Parker 1966, Mouschovias \etal 1974, Blitz \& Shu 1980, Elmegreen
1982b). \\
The latter application, however, has been criticized (Zweibel \& Kulsrud
1975, Parker 1975b, Balbus 1988), and recent observations indicate that
the Jeans instability is a major ingredient in this process (Elmegreen \etal
1994). It may be faster than the Parker instability, and self-gravity has
the advantage of not relying on the little known topology of the magnetic
field lines. The self gravitational collapse may even be helped by the
presence of a sheared magnetic field (Lynden-Bell 1966, Elmegreen \&
Elmegreen 1983, Elmegreen 1987). Some purely gravitational mechanisms have
also been advocated (Balbus 1988), as well as thermal instabilities
(Elmegreen 1989a,b, 1990). Nevertheless, the Parker mechanism may still play
a role (Elmegreen 1991, Hanawa \etal 1992), especially since recent work by
Giz \& Shu (1993) shows that, with more realistic vertical equilibrium
profiles, one should expect a growth rate a factor two higher than in
classical calculations. The Parker instability could also be a key
ingredient of the dynamo process in stars (Parker 1975a) and galaxies
(Ruzmaikin \etal 1988, Hanasz \& Lesch 1993  and references therein), if
combined with differential rotation.\\

The recent re-discovery of a powerful shearing instability (Velikhov 1959,
Chandrasekhar 1960) in magnetized discs by Balbus \& Hawley (1991) has
renewed the interest taken in differentially rotating magnetized discs: an
axisymmetric instability was found to occur in any magnetic configuration
which is not purely azimuthal (Balbus \& Hawley 1991, Dubrulle \& Knobloch
1993, Kumar \etal 1994). It could be efficient even in weakly ionized discs
(Blaes \& Balbus 1994).  A vertical magnetic field is also unstable to
non-axisymmetric perturbations (Tagger \etal 1992, 1993, 1995).\\

The configuration with a purely azimuthal magnetic field is stable to
axisymmetric perturbations, but transiently unstable to non-axisymmetric
ones. This different instability was very briefly mentioned by Acheson
(1978) in the context of stellar interiors, and described in more detail by
Foglizzo \& Tagger (1991), and Balbus \& Hawley (1992b) when applied to
galaxies and accretion discs.\\

One of the most promising results of these shearing instabilities, combined
with the Parker instability, could be the description, in a single model, of
both the turbulent viscosity and the dynamo action in accretion discs
(Vishniac \& Diamond 1992, Tout \& Pringle 1992). Unfortunately, these
attempts are still mainly qualitative, because of the analytical
difficulties introduced by differential rotation.\\

On the other hand, although the case of an initially vertical magnetic field
leads to the most powerful shearing instability, its nonlinear development
gives birth to a strong azimuthal magnetic component (Zhang \etal 1994).
This azimuthal magnetic field is subject to both the Parker and the
transient shearing instabilities, and has not been fully studied yet: Balbus
\& Hawley (1992b) canceled the effects of magnetic buoyancy by studying
the shearing instability in the disc midplane. Moreover, they introduced a
radial component of the initial magnetic field, thus precluding the
possibility of an initial hydrostatic equilibrium.\\

In this paper we describe in detail the effect of differential rotation on
waves in a disc embedded in an azimuthal magnetic field, at the linear
stage. This has already been the purpose of an earlier study (Foglizzo \&
Tagger 1994, hereafter [FT]), where we considered perturbations with a
vanishing vertical wavenumber because this is known to be the most
favourable configuration to the Parker instability, though it considerably
limits the effect of shear.\\
Here, by considering perturbations with a non vanishing vertical wavenumber,
acting on a true hydrostatic equilibrium (with a purely azimuthal initial
magnetic field), in a vertical density gradient, we can look at the details of
the conjugate effects of both the Parker and the azimuthal shearing
instability. We will show that they occur on the same branch (the ``slow
wave'') of the MHD dispersion relation, at different values of the radial
wavenumber. Since shearing motions lead the latter to vary with time, this
means that the same wave can evolve through periods of stable oscillations,
and shearing or Parker amplification, corresponding to a changing polarization
(\ie the orientation of the perturbed velocities associated with the wave),
respectively more radial or more vertical. These two amplification
mechanisms are found to act successively or sometimes simultaneously,
depending on the strength of differential rotation.\\

Our paper is organized as follows: in Sect.~2 we first recall our
hypotheses, which essentially reduce to the ``canonical'' ones used in
classical works on the Parker instability and to the use of the ``shearing
sheet'' model used in the description of spiral waves in differentially
rotating discs. This leads to our set of linearized differential equations.
In Sect.~3 we discuss in detail the validity of the WKB approximation, and
interpret it in terms of the natural polarization of the slow MHD waves.
Throughout the paper we will find that, in addition to giving a clear
understanding of the physics involved, the WKB approximation gives results
in excellent agreement with exact numerical solutions.\\
In Sect.~4 we explain the mechanism of the shearing instability. We compare
the characteristic features of the shearing and the Parker
instabilities, taken separately. We show analytically in Sect.~5 how these
two instabilities proceed from the same slow MHD mode, and solve the
apparent contradiction with the transient shearing stabilization found in
[FT]. We also derive the overall strength of the transient amplification.
\\
The phase of transition between the two instabilities is shown to depend
crucially on the strength of differential rotation in Sect.~6. We summarize
our main conclusions in Sect.~7, and give the details of
our analytical treatment in the appendices.

\section{The equations}
\subsection{Hydrostatic equilibrium}
The equilibrium is exactly the same as in [FT], and we recall it here very
briefly. In the rotating frame at distance $r_0$ from the rotation axis, we
define the vectors ${\bf x}$ along the radial direction ($x=r-r_0$), ${\bf
y}$ along the azimuthal magnetic field (${\bf B}_o=B_o {\bf y}$) and ${\bf
z}$ parallel to the rotation axis  (${\bf \Omega}=\Omega{\bf z}$).\\
The vertical gravitational field is taken to be constant ($ -{\rm sign}(z)g_z
{\bf z}$). This classical hypothesis allows an easier mathematical
formulation.\\
The magnetic pressure (and the cosmic ray pressure in the case of a galactic
disc) are assumed to be proportional to the thermal gas pressure:
\begin{equation}
P_{b}=\alpha P_{\rm th} \mbox{ and } P_{\rm cr}=\beta P_{\rm th}.
\end{equation}
The gas is convectively neutral ($\gamma=1$), and both the sound speed $a$
and the Alfven speed $\VA=a(2\alpha)^{1/2}$ are independent of
height.\\
Thus the vertical density profile of the hydrostatic equilibrium is simply:
\begin{equation}
\rho_o(z) = \rho_o(0) \exp(-|z|/H),
\end{equation}
where the scale height is:
\begin{equation}
H \equiv (1+\alpha +\beta){a^2\over g_z}.
\end{equation}
In our numerical calculations, we assume that the vertical gravity and
rotation frequency are related by:
\begin{equation}
g_z\sim {H\over r_0} g_r\sim H\Omega^2.
\label{rothaut}
\end{equation}
This neglects the gravity of the midplane disc of stars, and consequently
would  overestimate the scale height $H$ in the case of a realistic galactic
disc.

\subsection{Linearized perturbations in the shearing-sheet approximation}
We will write our set of equations in terms of the Lagrangian displacement
vector defined by:
\begin{equation}
({\p\over\p t}+ {\bf V}_o.{\bf\nabla}){\bf\xi}={\bf v}+ ({\bf\xi
.\nabla}) {\bf V}_o. \label{displacement}
\end{equation}
The perturbations of the density and magnetic field take the simple form:
\begin{eqnarray}
\rho&=&-\nabla .(\rho_o {\bf\xi}),\\
{\bf b}&=&{\bf\nabla}\times ({\bf\xi}\times {\bf B}_o).
\end{eqnarray}
In the shearing sheet approximation, differential rotation is measured by
the Oort constant $A$:
\begin{equation}
A\equiv {r_{0}\over 2}{\partial\Omega\over\partial r}(r_0)<0.
\end{equation}
In the frame rotating at radius $r_0$ with the rotation frequency
$\Omega=\Omega(r_0)$, the sheared motion is linearized so that the azimuthal
speed of the gas at equilibrium is:
\begin{equation}
V_0=r_0\Omega_0+2Ax.
\end{equation}
The only difference with [FT] is the presence here of a vertical wavenumber
$k_z$, so that perturbed quantities vary as:
\begin{eqnarray}
\xi(x,y,z,t)&=&\nonumber\\
\rho_o^{-{1\over2}}\e^{i(k_yy+k_zz)}& &\int_{-\infty}^{+\infty}
\e^{i(k_x-2Ak_yt)x} \tx(k_x,t)\d k_x\label{Fourierinv}
\end{eqnarray}
$k_z$ was set to zero in [FT] for the sake of simplicity, since classically
this is the most unstable Parker mode. In Eq.~(\ref{Fourierinv}) and hereafter
the ($k_y,k_z$)-dependence of $\tx(k_x,t)$ is not written explicitly, for the
sake of clarity, since linear calculations allow us to keep them constant.\\
Defining $\tx(k_x,t)$ by Eq.~(\ref{Fourierinv}) permits us to interpret one
effect of differential rotation as a linear time-dependence of the radial
wavenumber of the sheared perturbation ($K_x(t)=k_x-2Ak_yt$). This appears
on the differential system satisfied by $\tx(k_x,t)$:
\begin{equation}
{\p^2{\bf\tx}\over \p
t^2}=-2{\bf\Omega}\times{\p{\bf\tx}\over \p t}-{a^2\over H^2} {\cal
L}(K_x(t))[{\bf\tx}],\label{systdiff3}
\end{equation}
where the dimensionless hermitian matrix ${\cal L}(\bk_x)$ includes
additional terms due to the vertical wavenumber $k_z$:
\begin{eqnarray}
\left(
\begin{array}{ccc}
\displaystyle{\bk_x^2\!\!+\!\!2\alpha(\bk_x^2\!+\!\bk_y^2\!-\!\bk_A^2)}&
\displaystyle{\;\bk_x\bk_y\;}&
\displaystyle{({1\over 2}\!+\!\beta\!-\!i\bk_z(1\!+\!2\alpha))i\bk_x}\\ & &
\\ \times &
\displaystyle{\bk_y^2 }&
\displaystyle{({1\over2}\!+\!\alpha\!+\!\beta\!-\!i\bk_z)i\bk_y}\\ & & \\
\times & \times &
\displaystyle{2\alpha\bk_y^2\!+\!(\bk_z^2\!+\!{1\over4})(1\!+\!2\alpha)}
\end{array}
\right)     \nonumber
\end{eqnarray}
We have noted $\bk\equiv Hk$ the dimensionless wavenumbers normalized by the
scale height of the disc, and crosses denote the complex conjugate of the
symmetric terms.\\
The differential force appears through the dimensionless
parameter $\bk_A$ defined as:
\begin{equation}
\bk_A^2\equiv -{4A\Omega H^2\over \VAc}>0.
\label{defka}
\end{equation}
As stated in [FT], the shearing time-scale $T_{\rm Shear}=|A|^{-1}$ occurs
both:
\par (i) as the typical time-scale of the shearing of waves, \ie the linear
growth of the radial wavenumber:
\begin{equation}
K_x(t)\equiv k_x-2Ak_yt,
\label{klineaire}
\end{equation}
\par (ii) as a scaling of the radial differential force, which is opposite to
the radial magnetic tension. Their ratio is:
\begin{equation}
-{4A\Omega\over k_y^2\VAc}={k_A^2\over
k_y^2} \sim{2\pi\over\bk_y^2}{T_{\rm Parker}^2\over T_\Omega T_{\rm Shear}}.
\label{scaling}
\end{equation}

\section{WKB approximation}
\subsection{Mathematical transformations}
The presence of a vertical wavenumber $k_z$ still allows the mathematical
transformations performed in [FT]: we reduce the sixth order differential
system (\ref{systdiff3}) to a second order differential equation, by first
performing a Laplace transform:
\begin{equation}
\bx(k_x,\omega)=\e^{-{i\omega
k_x\over2Ak_y}} \int_0^{+\infty} \e^{i\omega t}\tx(k_x,t)\d t.
\label{Laplace}
\end{equation}
In Appendix~A we derive an ordinary differential equation, of which
$\bx(k_x,\omega)$ is the unique solution bounded at infinity. This equation
has a singularity where $\Delta(\omega^2)=0$, and this polynomial is
interpreted in [FT] as the asymptotic dispersion relation at
$|K_x(t)|\to\infty$.\\
Reciprocally, $\tx(k_x,t)$ is determined by the inverse Laplace transform:
\begin{equation}
\tx(k_x,t)={1\over
2\pi}\int_{ip-\infty}^{ip+\infty}\e^{-i\omega
\left(t-{k_x\over2Ak_y}\right)}\bx(k_x,\omega)\d\omega,
\label{invLaplace}
\end{equation}
where $p$ is real and larger than the largest imaginary part of the
singularities of $\bx$.\\ The solution of the differential equation
(\ref{Frobenius}) can be written in the WKB approximation:
\begin{eqnarray}
\tx(k_x,t)={1\over2\pi}\int_{ip-\infty}^{ip+\infty}
\left(\mu_+\e^{i\Psi_+}- \mu_-\e^{i\Psi_-}\right)\d \omega.
\label{saddle}
\end{eqnarray}
and the functions $\mu_\pm(\omega)$ vary slowly if a WKB criterion, derived
in Appendix~B, is satisfied. We show that the major contributions to the
integral (\ref{saddle}) are then given by the saddle points $\omega_\pm(t)$
defined by:
\begin{equation}
{\p\Psi_\pm\over\p\omega}(\omega_\pm(t))=0.
\label{WKBdisp2}
\end{equation}
As in [FT], we can move the integration contour of Eq.~(\ref{saddle}) so as
to pass through these saddle points, and follow the steepest descent path
between them. When the WKB criterion is satisfied, this formulation allows
us to interpret Eq.~(\ref{WKBdisp2}) as a dispersion relation whose roots
are time-dependent. The saddle points are the relevant {\it instantaneous}
frequencies or growth rates which used to be the poles of the Laplace
transform $\bx(\tom)$ in the absence of differential rotation.\\

\subsection{WKB validity and the linear coupling of waves}
It was shown in [FT] that if the WKB criterion is satisfied, the evolution is
``adiabatic", \ie a wave (stable or not) preserves its ``identity" throughout
its sheared evolution. On the other hand, at low $|K_x(t)|$, the WKB
approximation fails if the differential rotation is realistic, and a linear
coupling of waves occurs: an initial pure Parker mode (slow MHD wave) is
associated with a complex pattern of motions, due to the
interplay of the Coriolis force and magnetic tension with its basic
perturbed velocities. As a result it loses its identity and it can emit both
magnetosonic (fast MHD) and Alfvenic waves in the disc.

The WKB criterion can be understood as the requirement that the instantaneous
frequency and polarization of the waves evolve (because of the time
dependence of the radial wavenumber $K_x(t)$) sufficiently slowly with time.
This criterion is satisfied in two different cases (see Appendix B):

\par (i) If $K_x(t)$ varies very slowly with time, the very slow temporal
variations of the growth rate and polarization allow a globally adiabatic
evolution of the solution. This is made possible if the shearing parameter
$A$ is very small.\\
The effect of the differential force ($4A\Omega \xi_x$) becomes negligible
compared to the radial magnetic tension ($k_y^2\VAc \xi_x$) if the
magnetic field is not very weak (as in Shu 1974). Nevertheless, the
case of a vanishing field allows us to study the effect of the differential
force on the Parker instability ([FT] and here, Sect.~5).

\par (ii) If the polarization and the frequency of the perturbation depend
weakly on the radial wavenumber $K_x(t)$, the WKB analysis is valid even if
the shearing parameter is not small. We study in the next section the
polarization and the frequency of the slow MHD wave to determine the range
of wavenumbers $K_x,k_z$ that allows this approximation.

\subsection{The  polarization of the slow MHD mode}
The slow MHD mode is characterized by displacements with a strong component
along the magnetic field lines, taking advantage of the anisotropy of the
magnetic pressure. The amounts of radial and vertical displacements depend
crucially on the orientation of the wavevector. {\it Independently of the
presence of any instability}, the {\it vertical stratification} and the {\it
rotation} of the magnetized plasma are responsible for the following
properties:
\begin{eqnarray}
{\xi_z\over\xi_x}&=&{\cal O}\left({k_x\over k_z}\right)\to \infty\;\; {\rm
when~} |k_x|\gg k_z,Hk_y^2  \label{gkx}\\
{\xi_z\over\xi_x}&=&{\cal O}\left({k_x\over k_z}\right)\to 0 \;\;{\rm when~}
|k_z|\gg k_x,k_y^2a_s^2/\Omega  \label{gkz}\\
{\xi_z\over\xi_x}&\sim&-{k_x\over k_z}\;\;{\rm when~} |k_z|,|k_x|\gg Hk_y^2,
k_y^2a_s^2/\Omega   \label{gkzx}
\end{eqnarray}
The transition between these asymptotic polarizations occurs at $k_x\sim
k_z$ (if $k_z>Hk_y^2,k_y^2a_s^2/\Omega$). It should be remembered that in most
cases of interest for us, $Hk_y^2\sim k_y^2a_s^2/\Omega\sim 1/H$ \\
These properties are typical of both the Alfven and slow MHD modes in the
classical case of a stratified plasma with uniform rotation, and can be
derived in a straightforward way by a normal mode analysis (see Appendix~C).
The WKB approximation, which is valid in each of these asymptotic limits,
allows us to generalize these properties to the case of differential
rotation.\\

We know from [FT] that differential rotation causes a wave with an initial
radial wavenumber $K_x\sim-\infty$ to evolve linearly with time towards
$K_x\to+\infty$. A slow MHD wave with an initial $K_x(t=0)\ll -|k_z|$ is
polarized mostly in the vertical plane ($y,z$) (see Eq.~(\ref{gkx})). It rocks
naturally towards the horizontal ($x,y$) plane when, because of the shear,
$|K_x(t)|<k_z$ (see Eq.~(\ref{gkz})). It rocks again towards the vertical
($y,z$) plane when $K_x(t)>k_z$ (Eq.~(\ref{gkx}) again). \\
The intrinsic nature of the slow MHD wave is consequently enough to explain
that the displacements are successively vertical, horizontal, and vertical
again, and thus more prone to buoyancy or Coriolis forces. Figure~\ref{fpolar}
illustrates qualitatively this evolution.\\
\begin{figure}
\picture 86.7mm by 72.3mm (figpol)
\caption[]{Polarization of the slow magnetosonic mode during its sheared
evolution.}  \label{fpolar}
\end{figure}

When the radial wavenumber $|K_x|\gg k_z,Hk_y^2$, the polarization of the slow
MHD wave is essentially vertical and the eigenfrequency remains constant.
The dispersion relation  reduces to $\Delta(\omega^2)=0$, which is
independent of $k_x$. The evolution is then adiabatic and allows a WKB study
of the asymptotic Parker instability (Shu 1974, [FT]).\\

When the vertical wavenumber $|k_z|\gg k_x,k_y^2a_s^2/\Omega$, the
polarization is
essentially horizontal and the eigenfrequency remains constant. According to
Eq.~(\ref{klineaire}), the duration of the phase of horizontal polarization
scales as:
\begin{equation}
T_{\rm horiz.}\sim {k_z\over Ak_y}.
\label{hortime1}
\end{equation}
Thus the polarization of the slow MHD wave is constantly horizontal on an
arbitrary long time if the vertical wavenumber is high enough. The
dispersion relation reduces to $Q_A(\omega^2)=0$, which is independent of
$k_x$ (see Appendix~B). The evolution is then adiabatic and allows a WKB
approximation.\\

The strongest effect of the differential force is expected in this
phase of horizontal motions, since this force is radial and
proportional to the radial displacement. Indeed, if the differential force
dominates the radial magnetic tension, the shearing instability will result,
during a time comparable to $T_{\rm horiz.}$ (see Sect.~4-5).\\

In the opposite case where $k_z\le k_y^2a_s^2/\Omega\sim 1/H$, the phase of
horizontal polarization shrinks to about one shear time, and the evolution may
be adiabatic only in the limit of vanishing differential rotation (see [FT] for
$k_z=0$).\\

If both the vertical and radial wavenumbers are large $K_x\sim k_z\gg Hk_y^2,
k_y^2a_s^2/\Omega$, the polarization of the slow MHD wave varies slowly:
\begin{equation}
{\p K_x(t)/k_z\over \p t}\sim -2A{k_y\over k_z}\ll |A|,
\end{equation}
and the WKB approximation is possible again (see Sect.~6).

\section{The pure shearing instability}
\subsection{The pure shearing instability}
If a magnetic field line of length $\Lambda_y=2\pi/k_y$ is bent by a radial
displacement $\xi_x$, a magnetic tension ($F_{mt}\equiv\rho_0 k_y^2V_{\rm
A}^2\xi_x$) acts as a restoring force. The differential force
($F_d\equiv 4A\Omega \rho_0\xi_x$)
is also proportional to the radial displacement, but in the opposite
direction. Hence if the following criterion is satisfied:
\begin{equation}
4A\Omega+k_y^2\VAc<0,
\label{critshear}
\end{equation}
a radially bent field line will naturally continue its radial displacement.\\

This balance is indeed a central feature of the shearing instability.
However, this simple description is not enough to fully explain it: it
ignores the effects of the other forces, namely the Coriolis force, the
pressure
force and the other components of the magnetic force. In particular, the
Coriolis force is usually stabilizing by transforming any radial motion into an
epicyclic oscillation, as in the case of sound waves modified by rotation.\\
Moreover, one would like to understand why the criterion (\ref{critshear})
seems to predict an instability in the limit of a vanishing magnetic field,
which seems to contradict the stability ensured by the Rayleigh criterion
($\kappa^2=4\Omega^2+4A\Omega>0$) for a non-magnetized disc.

\begin{figure}
\picture 86.7mm by 72.3mm (figmech)
\caption[]{The shearing instability viewed from above the disc: the rotation
of the disc is here anti-clockwise. The amplitude of the perturbation is
exaggerated for the sake of clarity. The circles along the perturbed magnetic
field line represent the maxima of density. When the displacements of the slow
MHD mode are
mainly horizontal ($\lambda_z<\lambda_x$), the radial thermal pressure cancels
the anisotropic magnetic pressure. Instability occurs if the radial magnetic
tension is dominated by the differential force. The azimuthal thermal
pressure force acts as a restoring force, which can overcome the stabilizing
influence of the Coriolis force.} \label{fmechshear}
\end{figure}

We now proceed to a detailed description of the mechanism, by analysing the
linearized system (\ref{systdiff3}) in the ``horizontal regime" where $k_z\gg
1/H, k_x$, and Eq.~(\ref{gkz}) is valid. In particular the stratification and
cosmic ray effects disappear here. \\
An illustrative picture of the mechanism is given in Fig.~\ref{fmechshear}.\\

It is convenient, for the sake of clarity, to decompose the magnetic force into
two contributions, which have simple expressions in our approximation:
\par (i) an anisotropic ``pressure" force acting against the compression of the
field lines,
\begin{eqnarray}
{\bf F}_{pb}\equiv i\rho_0\VAc {\bf\nabla}_\perp(\nabla . {\bf \xi}_\perp)
=-i\rho_0\VAc(K_x\xi_x+k_z\xi_z)
\left\{
\begin{array}{l}
 K_x\\0\\k_z
\end{array}
\right.
\end{eqnarray}
\par (ii) a magnetic tension which can be visualized as the result of the
``elasticity" of the field lines when they are bent.
\begin{eqnarray}
{\bf F}_{tb}\equiv -i\rho_0\VAc k_y^2 {\bf \xi}_\perp=-i\rho_0\VAc k_y^2
\left\{
\begin{array}{l}
 \xi_x\\0\\ \xi_z
\end{array}
\right.
\end{eqnarray}

The $z-$component of Eq.~(\ref{systdiff3}) implies that in a slow MHD mode in
the horizontal regime, {\it the thermal pressure force cancels the magnetic
anisotropic pressure force}, to first order:
\begin{equation}
\VAc\nabla .({\bf \xi}_\perp) +a_s^2\nabla . {\bf \xi}\sim 0 \label{denspert}
\end{equation}
Note that this does {\it not} mean incompressibility: Eq.~(\ref{denspert})
can be equivalently written in order to express the density perturbation as a
function of the azimuthal displacement:
\begin{equation}
{\rho\over\rho_0}=-{i k_y\VAc\over a_s^2+\VAc}\xi_y\label{denspert2}
\end{equation}
Independently of the presence of any instability, this implies in
particular that the extrema of the density perturbed by a slow MHD mode occur
where the azimuthal displacement vanishes (see Fig.~\ref{fmechshear}).\\

\par(i) As a consequence of Eq.~(\ref{denspert}), the sum of the thermal
pressure force and the magnetic force, in the radial direction, is reduced to
the single magnetic tension term mentioned above.
\par(ii) As a consequence of Eq.~(\ref{denspert2}), the azimuthal component of
the thermal pressure force can be expressed as a restoring force acting on the
azimuthal displacement.\\

Hence in the limit of small vertical wavelength, the slow MHD mode
(the Alfven mode as well) corresponds to perturbations satisfying the following
2-D system:
\begin{eqnarray}
\left\{
\begin{array}{l}
\displaystyle{{\p^2\over\p t^2}\xi_x=2\Omega{\p\over\p t}\xi_y
-(4A\Omega+k_y^2V_{\rm A}^2)\xi_x ,}\\
\displaystyle{{\p^2\over\p t^2}\xi_y=-2\Omega{\p\over\p t}\xi_x-
k_y^2V_c^2\xi_y.}  \end{array}
\right.
\label{syst2d}
\end{eqnarray}
The only remaining forces are the Coriolis force, the radial magnetic
tension modified by the differential force, and an azimuthal restoring force
(we have noted $V_c\equiv a_s\VA/(a_s^2+\VAc)^{1/2}$ the cusp speed).\\
We can
notice the similarity with the case of the axisymmetric instability of a disc
embedded in a magnetic field which is not purely azimuthal: Balbus \& Hawley
(1992a) obtained an equation similar to Eq.~(\ref{syst2d}), with the following
differences:   \par (i) $k_z$ appears instead of $k_y$ here, because of the
different geometry of the equilibrium magnetic field,
\par(ii) the azimuthal restoring force involves the
Alfven speed instead of the cusp speed here.\\
They stressed the link with the Hill equations, and gave a simple
mechanical analogue: the motion of a particle attached with a spring to
a guiding center in orbit in a gravitational potential. The spring
is isotropic ($C_x=C_y=k_z^2\VAc$). This analogy also holds in our case, by
considering an anisotropic spring ($C_x=k_y^2\VAc,C_y=k_y^2V_c^2$). One can be
easily convinced that {\it the motion of the particle is unstable if the
stabilizing effect of the spring is canceled by the differential force}
($4A\Omega+C_x<0$).\\
The importance of the azimuthal restoring force in
the instability mechanism is now clear: the Coriolis force would be able to
reduce the shearing instability to a neutral mode if the azimuthal spring
constant $C_y$ were to vanish (for example if the sound speed
$a_s\to 0$).\\
 This occurrence of the cusp speed may be traced back to the fact that we are
here dealing with the destabilization of the slow MHD mode, whereas the
axisymmetric instability studied by Balbus and Hawley corresponds to the
destabilization of the Alfven mode.\\

As shown in Appendix~B, the WKB criterion is satisfied in the horizontal
regime, so that we can replace time derivatives in Eq.~(\ref{syst2d}) by simple
($-i\omega$) multiplications (note that radial derivatives have disappeared).
The resulting dispersion relation reduces to $Q_A(\omega^2)=0$:
\begin{equation}
\omega^4\!-\![\kappa^2\!+\!2(1\!+\!\alpha)k_y^2V_c^2]\omega^2+
k_y^2V_c^2(4A\Omega\!+\!k_y^2\VAc)=0
\label{QA}
\end{equation}
According to this dispersion relation, instability indeed occurs if the
differential force dominates the magnetic tension.\\
The other root of this dispersion relation, corresponding to the Alfven
mode, is always stable and reduces to an epicyclic oscillation (inertial wave)
in the case of a vanishing magnetic field.\\
The missing mode is of course the fast magnetosonic mode, corresponding to
fast vertical oscillations, which was excluded from our approximation by
Eq.~(\ref{gkz}).\\

As the Parker instability, the shearing instability requires a long
enough wavelength along the magnetic field lines: in the galactic case, the
minimum azimuthal wavelength scales as $\pi\VA |A\Omega|^{-1/2}\sim
1.5$kpc.\\

The optimal azimuthal wavenumber corresponds to
$k_y^2\VAc=(1+\alpha)\omega_{\rm s}^2-2A\Omega$, where $\omega_{\rm s}$ is the
optimal growth rate:
\begin{equation}
\omega_{\rm s}^2={-2A^2\over
(1+2\alpha)^{1\over2}\left(1+2\alpha+2\alpha{A\over\Omega}\right)^{1\over2}
+1+2\alpha+{A\over\Omega}\alpha}.
\end{equation}
$|\omega_{\rm s}|$ is a decreasing function of the intensity of the magnetic
field $\alpha$. Thus the highest growth rate corresponds to a vanishing
magnetic field, and we obtain in this limit, for an infinite optimal $k_y$:
\begin{equation}
\omega_{\rm s}^2\to -A^2.
\end{equation}
This extreme limit is of course never reached, because the diffusion of the
magnetic field must be taken into account as soon as the magnetic field is
weak: our idealized MHD equations then break down, as shown in details by
Acheson (1978).\\
Nevertheless, this calculation confirms the study by Balbus \& Hawley
(1992a) suggesting that the Oort constant $A$ is the maximum growth rate for
such shearing instabilities.\\

The instability still exists when the magnetic field is strong, but the
growth rate decreases as $\alpha^{-1/2}$. This instability may play an
important role, not only in weakly magnetized accretion discs, but also in
the gaseous disc of galaxies, where the magnetic pressure is comparable to
the thermal pressure: if $\alpha\sim 1$ and $A/\Omega=-0.5$, the e-folding time
of the shearing instability is about one half of the rotation period for a
$m=1$ perturbation at $R\sim 10$kpc.\\

We can notice here an important difference with the axisymmetric instability of
a disc embedded in vertical magnetic field: in this latter case, the
instability criterion again compares the differential force to the magnetic
tension: $4A\Omega+k_z^2\VAc<0$. In particular, this condition
implies that the instability cannot exist if the magnetic
energy is comparable to or larger than the thermal energy. Indeed, the vertical
wavelength of the perturbation must be shorter than the scaleheight of the
disc, so that axisymmetric instability requires (Balbus \& Hawley 1991):
\begin{equation}
\VA<{1\over \pi}|A\Omega|^{1\over2}H\sim {\rm Max}(a_s,\VA).
\end{equation}
By contrast, the azimuthal wavelength is only bounded by the circumference of
the disc, so that the non-axisymmetric instability we are studying occurs as
long as:
\begin{equation}
\VA<{1\over \pi}|A\Omega|^{1\over2}R\sim {\rm Max}(a_s,\VA){R\over H}.
\end{equation}
The shearing instability might therefore be efficient for magnetic fields
exceeding the equipartition value, in the outer parts of the disc. This might
be an important property of the instability, {\it especially if it were related
to a dynamo mechanism}.\\
\begin{figure}
\picture 86.7mm by 72.3mm (1a)
\picture 86.7mm by 72.3mm (1b)
\caption[]{With  $k_A>k_y>k_{\rm P}$, the perturbation is stable against the
Parker instability at $|K_x(t)|>|k_z|$, and unstable to the shearing
instability at $|K_x(t)|<|k_z|$. The evolution of the displacement vector
(top figure), is in very good agreement with the calculation of the slow MHD
root of the WKB-dispersion relation (bottom figure). Parameters are
$-A/\Omega=.5, k_z=10., \alpha=0.1, \beta=0$. The three components
($x,y,z$) of the displacement vector $\xi$ are displayed. Wavenumbers are in
units of $H^{-1}$, frequencies in units of $V_{\rm A}/H$.}
\label{PureShearing}
\end{figure}

\subsection{Comparison of the growth rates}
It is interesting at this point to compare the characteristic length scales
and growth rates of the optimal shearing and Parker instabilities. Both
occur at large azimuthal scales, \ie at low $k_y$. The Parker instability
requires $k_y<k_{\rm P}$ at $|k_x|\to\infty$ (where $k_{\rm P}$ is defined in
Eq.~(\ref{defkp})), whereas the shearing instability requires  $k_y<k_A$ at
$|k_z|\to\infty$ (where $k_A$ is defined in Eq.~(\ref{defka})).\\
As an illustration,
Fig.~\ref{PureShearing} (top) shows the shearing instability occurring at an
azimuthal wavelength which is stable against the Parker instability
($k_A>k_y>k_{\rm P}$). It was obtained by integrating numerically the
differential system (\ref{systdiff3}). The bottom graph shows the corresponding
slow-wave solutions of the WKB dispersion relation. The two solutions
(corresponding to slow waves propagating in opposite directions) are not
exactly symmetric with respect to the $\omega$ axis, showing an effect of
rotation which was already noted by Shu (1974). The two solutions merge at
$K_x\simeq -k_z$ where they become unstable. The numerical solution shows that,
as a result, when one starts with a ``pure'' solution on the top branch, one
gets at positive $K_x$ a mixture of the two waves. This illustrates the linear
coupling, which occurs because the WKB criterion is not strictly satisfied
where the two slow MHD waves merge: the change of polarization and frequency
between the oscillation phase and the growing phase is too fast (the shear,
corresponding to a flat rotation curve, is strong: $-A/\Omega=.5$).
Figure~\ref{PureShearing}b illustrates this rapid change.\\

We have plotted in Fig.~\ref{SVersusP} the range of magnetic pressures and
rotation profiles for which, for any azimuthal wavenumber, the shearing
growth rate is always higher than the Parker growth rate (contour levels
$>1$), and conversely (contour levels  $<1$). For instance, along the
contour level labeled ``2", the shearing growth rate is twice as large as
the optimal Parker growth rate, at the wavenumber $k_y$ which is optimal for
the Parker instability.\\
In the transition between these two regions, the choice of the azimuthal
wavenumber determines which instability is faster. Introducing a cosmic-ray
gas favours the Parker instability only: the pure-shearing instability does
not depend on the cosmic-ray pressure, because the displacements involved
are essentially horizontal and the differential force does not act on the
massless cosmic-ray gas (see Eq.~(\ref{QA})).\\

A ``realistic'' galactic disc ($\alpha\sim\beta\sim1,\;\; -A/\Omega\sim .5\;
{\rm to }\; .75$) would lie in the zone where the strengths of the two
instabilities are comparable. On the basis of the comparison, in the linear
regime, of the growth rates of these two instabilities, {\it it is not possible
to favour one particular vertical wavenumber}. This contrasts with the case
of a uniformly rotating disc, where the maximization of the growth rate
of the Parker instability over all possible wavenumbers would select the
mode with $k_z=0$ (Parker 1966, Zweibel \& Kulsrud 1975).\\
Non-linear effects will play a major role in selecting the vertical wavelength
of the most unstable perturbations, but this is beyond the scope of the
present paper. \\

In an accretion disc with Keplerian rotation ($-A/\Omega\sim .75$), the
growth rate of the shearing instability is higher than the Parker growth
rate at any $\alpha<2$. In such a disc, the fastest growing perturbations, in
the linear regime, have a short vertical wavelength and are essentially
horizontal.\\

We must bear in mind that this comparison is quantitatively dependent on the
relationship assumed between the scale height $H$ and the rotation frequency
$\Omega$ (Eq.~(\ref{rothaut})). Figure~\ref{SVersusP} is, however,
qualitatively relevant. A lower scale height would slightly favour the
Parker instability. On the other hand, if the scaleheight of the magnetic
pressure were higher than the scale height of the gas, the Parker instability
would be weakened, while the shearing instability would remain unchanged.\\
\begin{figure}
\picture 86.7mm by 72.3mm (SVersusP)
\caption[]{Comparison of the growth rates of the shearing and Parker
instabilities, depending on the strengths of the magnetic field and
differential rotation. The contour levels give a conservative estimation of
the ratio $\omega_{\rm S}/\omega_{\rm P}$: it is minimized (resp. maximized)
over every $k_y$ when $\omega_{\rm S}/\omega_{\rm P}>1$ (resp. $\omega_{\rm
S}/\omega_{\rm P}<1$).}
\label{SVersusP}
\end{figure}

This comparison of the two isolated optimal instabilities is of course only a
first estimate of their relative importance. A refined study would require
taking into account the temporal evolution of the radial wavenumber, and
noting how the instabilities may be simultaneous or mutually exclusive.\\
We already know that the Parker instability is favoured by large radial
wavenumbers and vanishing vertical ones, whereas the shearing instability
favours large vertical wavenumbers and limited radial ones. The next
sections are devoted to a further study of this point.

\section{The Parker-Shearing instability}
\subsection{The simplified dispersion relation for slow MHD waves}
As mentioned by Balbus \& Hawley (1994), the shearing instability is the
consequence of the differential force rather than of the shearing of the
flow. In particular, this local instability would not exist in a shear flow
between two parallel planes, as can be easily checked
by canceling the rotation terms (Coriolis and differential forces) and keeping
the shearing of waves (time-dependent wavenumber).\\

Nevertheless, the sheared
motion is responsible for the transience of the instability, as well as a
linear  coupling of the waves at low $|K_x(t)|$, as was shown in [FT]. We
concentrate here on the nature of the Parker and shearing instabilities, and
avoid the complications due to this coupling by considering the case of low
shear ($A/\Omega\to 0$).\\

In order to study the effects of the differential force, we keep it
comparable with the magnetic tension, \ie
\begin{equation}
k_y^2\VAc\sim 4A\Omega.
\end{equation}
This leads us to consider the case of a vanishing magnetic field ($\alpha\to
0$) so as to keep:
\begin{eqnarray}
k_A & \sim & k_y,\\
\tilde\Omega^2& \sim & 1/\alpha\gg 1.
\end{eqnarray}
Hence we use the same WKB-approximation as in [FT], but include here a
non-vanishing vertical wavenumber $k_z$. Assuming that $\tom={\cal
O}(1/\tilde\Omega)$, we can write the leading terms of  the WKB-dispersion
relation (\ref{dispers}) and obtain:
\begin{equation}
\left(\bk_z^2+{1\over4}\right)(\tom\tilde\Omega)^2+2\bar K_x\bk_y\bk_z
(\tilde\Omega\tom)-\bk_y^2R=0,
\label{frequency}
\end{equation}
with:
\begin{equation}
R= \bk_z^2(\bk_y^2-\bk_{\rm A}^2)+
\bk_y^2(\bk_y^2-\bk_{\rm Q}^2)+\bk_{\rm A}^2/4-\bar K_x^2(\bk_{\rm P}^2-
\bk_y^2).
\end{equation}
When $k_A=0$, we recover the classical criterion for Parker instability at
$K_x=0$:
\begin{equation}
k_y^2+k_z^2<k_{\rm Q}^2.
\end{equation}
This indicates that the dispersion relation (\ref{frequency}) gives, for
finite values of $K_x(t)$, the behaviour of the two slow MHD frequencies.\\

Now we study the stabilizing or destabilizing influence of the differential
force on the slow MHD branch.

\subsection{Stabilization of the Parker instability by differential rotation}
When $\bk_z^2<1/4$, a strong differential force ($k_A>k_y$) impedes the
Parker instability at low $|K_x(t)|$ even though $k_y^2+k_z^2<k_{\rm Q}^2$.
This stabilization was found in [FT] for $\bk_z=0$, and is illustrated in
Fig.~\ref{Lowkz} for $\bk_z=0.25$. Nevertheless, this stabilization lasts a
few shear times only.
\begin{figure}
\picture 86.7mm by 72.3mm (2a)
\picture 86.7mm by 72.3mm (2b)
\caption[]{Transiently stabilized Parker instability. The shearing
instability criterion is satisfied ($k_y<k_A$), the differential rotation is
stabilizing because $\bk_z<1/2$. Parameters are $\alpha=.4$, $\beta=0$,
$-A/\Omega=.1$, $\bk_y=\bk_z=0.25$. Consequently, $\bk_{\rm Q}=0.67$, and
$\bk_A\sim \bk_{\rm P}=0.84$. Here, the stabilization lasts about $4$ shear
times}
\label{Lowkz}
\end{figure}

\subsection{Destabilization of the slow MHD mode by differential rotation}
When $\bk_z^2>1/4$, the discriminant of Eq.~(\ref{frequency}) is negative
if  $\bar K_x(t)^2<\bar K_s^2$, with:
\begin{equation}
\bar K_s^2\equiv (4\bk_z^2+1){\bk_y^2(\bk_{\rm Q}^2-
\bk_y^2-\bk_z^2)+\bk_A^2(\bk_z^2-1/4) \over
4\bk_z^2(\bk_y^2+1/2)+\bk_y^2-1/2}.
\label{formuleKo}
\end{equation}
The differential term $k_A^2$ clearly contributes to destabilize the slow MHD
mode, resulting in the shearing instability.\\

When $\bk_z\gg \bk_y,\bar K_x,1/2$, we recover the pure shearing instability
studied in Sect.~4:
\begin{equation}
\tom_{\rm S}^2\sim -{\bk_y^2\over\tilde\Omega}(\bk_A^2-\bk_y^2),
\end{equation}
with the usual criterion for instability:
\begin{equation}
k_A>k_y \Longleftrightarrow 4A\Omega+k_y^2\VAc<0.
\end{equation}
For $\bk_y^2=\bk_A^2/2$ we recover the optimal growth rate:
\begin{equation}
|\omega_{\rm S}|\sim |A|.
\end{equation}

\subsection{Efficiency of the shearing instability at $\bk_z>1/2$}
Eq.~(\ref{frequency}) is useful in determining the overall strength of the
transient shearing amplification. When the discriminant of
Eq.~(\ref{frequency}) is negative, the complex root is:
\begin{equation}
\tilde\Omega\tom=-{\bk_y\bk_z\bar K_x\over \bk_z^2+1/4}\lbrack 1\pm i
\left\{1+\left(\bk_z^2+{1\over 4}\right) {R\over\bk_z^2\bar
K_x^2}\right\}^{1\over2} \rbrack.
\end{equation}
We define the complex amplification factor $\varphi$ so that the amplitude of
the wave after the transient amplification is multiplied by $\exp(\varphi)$:
\begin{equation}
\varphi\equiv\int i\omega(t) \d t={\VA\over 2A\bk_yH}\int_{-\bar
K_s}^{+\bar K_s}   i\tom(\bar K_x) \d \bar K_x .
\end{equation}
The imaginary part of $\varphi$ vanishes because Im($\tom$) is odd in $\bar
K_x$, and we obtain:
\begin{equation}
\varphi=\pm{\pi\over \bk_A^2}{\bk_y^2(\bk_{\rm
Q}^2-\bk_y^2-\bk_z^2)+\bk_A^2(\bk_z^2-1/4)\over
\left(4\bk_z^2(\bk_y^2+1/2)+\bk_y^2-1/2\right)^{1/2}}.
\end{equation}
The amplification factor is an unbounded increasing function of $\bk_z$. For
large values of $\bk_z$, it scales as:
\begin{equation}
\varphi\sim \pm{\pi\over 2}{\bk_z\over
(\bk_y^2+1/2)^{1/2}}{\bk_A^2-\bk_y^2\over \bk_A^2}.
\end{equation}
The transient destabilization may be arbitrarily long if the vertical
wavenumber $k_z$ is large enough.\\

\subsection{Comment on the transient character of the Shearing instability}
We wish to emphasize that the shearing instability, although transient and
occurring preferentially at low $k_x$, is essentially {\it local} and does grow
on a sufficiently short time scale that its transient nature poses no physical
limit to its growth: this contrasts with \eg spiral density waves, which travel
in the disc and can experience amplification only for a limited time (as they
are reflected from corotation), unless the boundary conditions at the disc
center allow them to be reflected back toward corotation.\\

Indeed the time required for the growing perturbation to travel, vertically
or radially, along a distance comparable to the scaleheight of the disc is much
longer than the typical e-folding time for realistic values of differential
rotation. This can be checked by derivating the dispersion relation
(\ref{dispgkz}) with respect to $k_x$ (or $k_z$) to obtain the radial (or
vertical) group velocity: \begin{equation}
{v_{gx}\over H}={1\over H}{\p\omega\over\p k_x}\sim {\Omega\over Hk_z}\ll |A|
\end{equation}
Thus perturbations can be considered to grow where they were created.\\

The transient character of any instability may be an obstacle to its
development only if the instability does not have enough e-folding time to
produce a significant effect: we have shown that the shearing instability does
not suffer this limitation if we allow for a short enough vertical
wavelength.\\

Incidentally, it has been shown in the context of non-magnetized discs that
any local instability is stabilized by the effects of viscosity in a
differentially rotating disc (see  Korycansky (1992) or Dubrulle \&
Knobloch (1992)). This stabilization arises when the sheared radial wavenumber
increases towards infinity, \ie when the radial wavelength is smaller than the
typical scale for dissipative processes. Nevertheless, such a mathematical
limitation of the linear regime has no physical implication if the ``transient"
instability has had enough e-folding time to produce perturbations which lead
to the non linear regime.\\

As an illustration, we know that the Parker instability is stabilized by
ambipolar diffusion for scales $<1pc$ in the galaxy (see Cesarsky 1980): this
leaves more than 100 $e-$folding time for the Parker instability to occur on
larger radial scales.\\

Similarly, the shearing instability is limited by the minimum vertical scale
at which the ideal MHD equations are valid. It can be estimated, in a
galactic disc, as the same critical scale used in the radial direction for the
Parker instability. Although transient {\it by definition}, the shearing
instability is not {\it physically} more transient than the Parker instability.

\begin{figure}
\picture 86.7mm by 72.3mm (3a)
\picture 86.7mm by 72.3mm (3b)
\caption[]{Shearing instability at $|K_x|<K_s\sim 20$, and Parker
instability at $|K_x|>K_p\sim 40$, separated by an oscillating phase. Here
the shearing growth rate is lower than the Parker one. Parameters are
$\alpha=1$, $\beta=0$, $-A/\Omega=.25$, and $\bk_y=.57$ is optimal for the
Parker  instability $\bk_z=20$. We have then $\bk_A=\bk_{\rm P}=1.$ and
$\bk_{\rm Q}=.87$ }
\label{DomParker}
\end{figure}
\begin{figure}
\picture 86.7mm by 72.3mm (4a)
\picture 86.7mm by 72.3mm (4b)
\caption[]{Shearing instability at $|K_x|<K_s\sim 25$, and Parker
instability at $|K_x|>K_p\sim 65$, separated by an oscillating phase. Here
the shearing growth rate is larger than the Parker one. Parameters are
$\alpha=1$, $\beta=0$, $-A/\Omega=.5$, $\bk_z=20$, and $\bk_y=.89$ is
optimal for the shearing instability. Then $\bk_A=1.41$, $\bk_{\rm P}=1.$,
and $\bk_{\rm Q}=.87$ }
\label{DomShearing}
\end{figure}

\section{Transition between the Parker and the shearing instability}
\subsection{General evolution of the unstable slow MHD wave}
The dispersion relation (\ref{frequency}) demonstrates that both the
shearing and the Parker instabilities belong to the same slow MHD branch,
thus confirming the qualitative argument of Sect.~3.3.: the Parker
instability occurs when the polarization of the slow MHD wave is vertical
($K_x(t)\ll -|k_z|$), and is stabilized when the polarization rocks towards
the plane of the disc ($|K_x(t)|<k_z$). The differential force then acts, in
the horizontal plane, on the radially displaced gas in the same
destabilizing way as did the gradient of magnetic pressure, in the vertical
plane, on the vertically displaced gas. While the shearing goes on, the
polarization rocks back towards the vertical plane, leading to the Parker
instability again at $K_x(t)\gg k_z$. This behaviour is illustrated in
Figs.(5--8).\\

The relative growth rates of the two successive instabilities
(Parker/shearing) simply depend on the strengths of the magnetic field and
differential rotation, as studied in Sect.~4. The two instabilities can
follow one another without any interruption of growth as $K_x(t)$ varies
(Fig.~\ref{sansos}), or they can be separated by a phase where the slow MHD
mode is oscillating  as in Figs.(6--7).

\subsection{The case of weak differential rotation}
Even without differential rotation, the Parker instability is naturally
stabilized at low $k_x$ when the vertical wavenumber is high. This appears
on the classical criterion for the Parker instability $k_y^2+k_z^2<k_{\rm
Q}^2$ at $k_x=0$. This defines a radial wavenumber $K_p$ such that
stabilization occurs at $|k_x|<K_p$. It can be calculated from the classical
dispersion relation without differential rotation (\ref{classicdisp}). An
easy estimate of $K_p$ is possible in the absence  of rotation
($\tilde\Omega=\bk_A=0$), for large values of $\bk_z$:
\begin{equation}
\bar K_p\sim\left({\bk_y^2\over \bk_{\rm P}^2-\bk_y^2}\right)^{1/2}\bk_z.
\end{equation}

The shearing instability occurs at low $|K_x(t)|$ only, and is stabilized
for $|K_x(t)|>K_s$ even when the Parker instability is not present (see
Balbus \& Hawley 1992b). The value of $K_s$ depends on the intensity of the
differential rotation as in Eq.~(\ref{formuleKo}): $K_s^2$ is negative if
$k_A<k_y$, and it naturally increases when the shear parameter increases. We
obtain, for $\bk_z\gg\bk_y$:
\begin{equation}
\bar K_s\sim\left({\bk_A^2-\bk_y^2\over\bk_y^2}\right)^{1/2}\bk_z.
\end{equation}

A weak differential rotation is therefore likely to imply $K_s<K_p$.

\subsection{The case of realistic differential rotation}
\begin{figure}
\picture 86.7mm by 72.3mm (5a)
\picture 86.7mm by 72.3mm (5b)
\caption[]{Parker-Shearing instabilities occurring successively: we can
clearly see the two instabilities occurring at the same time while
$K_s>|K_x(t)|>K_p$, with $K_s\sim 50$ and  $K_p\sim 20$; this appears as a
distinct bump in the instant growth rate (bottom figure). The shearing
growth rate is slightly larger than the Parker one. Parameters are
$\alpha=1$, $\beta=0$, $-A/\Omega=.75$, $\bk_z=20$, and $\bk_y=.57$ is
optimal for the Parker instability. Then $\bk_A=1.73$, $\bk_{\rm P}=1$., and
$\bk_{\rm Q}=.87$ }
\label{sansos}
\end{figure}
We can derive a criterion for the presence of an oscillating phase of
transition between the two instabilities, for large values of $\bk_z$.
According to our previous estimate, such a transition occurs at $\bar
K_x\sim \bk_z$. The WKB-approximation is then possible even if the
differential rotation is strong (see Appendix~B).\\

If an oscillating transition occurs, the two slow MHD frequencies $\tom(\bar
K_x)$ are purely real, and satisfy the following dispersion relation:
\begin{equation}
Q_A+\left({\bar K_x\over\bk_z}\right)^2\Delta-2
\tilde\Omega\tom(1+\alpha+\beta)\bk_y \left({\bar K_x\over\bk_z}\right)=0.
\end{equation}
The two frequencies $\tom(K_x)$ pass through an extremum (see
Fig.~\ref{DomParker} and Fig.~\ref{DomShearing}) at $\tom_\pm(K_\pm)$ such
that:
\begin{equation}
{\p\tom\over\p\bk_x}=0\Longleftrightarrow {\bar K_\pm\over\bk_z}={\bar
\Omega\tom_\pm(1+\alpha+\beta)\bk_y \over\Delta}.
\end{equation}

In the limit of $k_z\gg1$, the two extremal frequencies $\tom_\pm$ are thus
solutions of the polynomial of degree 4 in $\tom^2$:
\begin{equation}
\Delta Q_A=\tilde\Omega^2\tom^2(1+\alpha+\beta)^2\bk_y^2.
\label{translim}
\end{equation}
The limiting case of a vanishing period of oscillations corresponds to
$\tom_+=\tom_-$ and $K_+=K_-$. This condition means that the polynomial
(\ref{translim}) has a double root. It can be translated into a condition on
its coefficients, depending on $\bk_y,\alpha,\beta,A/\Omega$. It is solved
numerically and displayed on Fig.~\ref{figtrans}. We can stress the
following properties, which are independent of $(\alpha,\beta)$:
\begin{figure}
\picture 86.7mm by 72.3mm (figtransa)
\picture 86.7mm by 72.3mm (figtransb)
\caption[]{Range of azimuthal wavenumbers leading to an oscillating
transition between the instabilities ($K_s<K_p$), or a phase of simultaneous
instabilities ($K_s>K_p$), depending on the strengths of the magnetic field
and differential rotation. The dotted line corresponds to $K_s=K_p$. In the
top figure, $\beta=0$ and $\alpha=1$. At $-A/\Omega=0.5$ (bottom figure),
the two instabilities are never simultaneous. The two dotted curves
correspond to the two critical wavenumbers $k_y$ leading, for any strength
$\alpha$ of the magnetic field, to $K_s=K_p$. Any other $k_y$ implies
$K_s<K_p$. }
\label{figtrans}
\end{figure}
\par (i) If the rotation curve decreases more slowly than the flat rotation
curve ($-A/\Omega<.5$), the two instabilities will always be separated by an
oscillating transition ($K_s\le K_p$). If $-A/\Omega$ is close enough to the
critical value (.5), the duration of the transition can be arbitrarily
short: there is a unique azimuthal wavenumber $k_y$ leading to $K_s=K_p$ .
Even in this case, the transition is slow because the growth rate still
vanishes between the two instabilities (see Fig.~\ref{finstant} as an
illustration).
\begin{figure}
\picture 86.7mm by 72.3mm (6a)
\picture 86.7mm by 72.3mm (6b)
\caption[]{Optimal transition between the Parker and the shearing
instabilities when $-A/\Omega<.5$: the growth rate vanishes between the two
instabilities. Here, $\alpha=1$, $\beta=0$, $-A/\Omega=.4$, $\bk_z=20$, and
we choose $\bk_y=.48$ according to Fig.~\ref{figtrans} so that
$K_s=K_p(\sim 35)$. Then $\bk_A=1.26$, $\bk_{\rm P}=1.$, and $\bk_{\rm
Q}=.87$}
\label{finstant}
\end{figure}
\par (ii) If the rotation curve decreases faster than the flat rotation
curve ($-A/\Omega>.5$), the two instabilities can follow each other
continuously, and even be simultaneous: $K_s> K_p$ as  in Fig.~\ref{sansos}
where the rotation is Keplerian ($-A/\Omega=.75$).
\par (iii) If the rotation curve is exactly flat ($-A/\Omega=.5$),
Fig.~\ref{figtrans} (bottom) shows that there are two critical values of
$k_y$ allowing $K_s=K_p$, for any strength of the magnetic field. This
demonstrates that the lowest value of $-A/\Omega$ permitting $K_s>K_p$ is
precisely $-A/\Omega=.5$, for any $\alpha$.

\section{Discussion and conclusions}
Our study reveals that the WKB approximation is a very powerful tool to
study analytically the effect of differential rotation on the slow MHD
waves. Our numerical computations are in very good agreement with the WKB
dispersion relation, even when the WKB criterion is not strictly satisfied.
We summarize here our main conclusions:\\

\par (i) The shearing instability comes from the non-axisymmetric
destabilization of the slow MHD wave by the differential force. However, a
parallel can be drawn with the axisymmetric shearing instability of the Alfven
mode in a disc embedded in a vertical magnetic field.

\par (ii) The Parker instability and the shearing instability both
correspond to the destabilization of the {\it same} slow MHD mode. The
occurrence of each instability can be understood by looking at the natural
polarization of the slow MHD mode, depending on whether their radial
wavelength is longer or shorter than their vertical wavelength.

\par (iii) Although transient, the shearing instability may last an
arbitrarily long time if the vertical wavelength is short enough. In this
respect it is a true local instability. Moreover, it still exists for a
magnetic field strength exceeding the equipartition value.

\par (iv) The action of differential rotation is opposite depending on
whether the vertical wavelength of the slow MHD wave is longer or shorter
than the scale height of the disc. Perturbations with a long vertical
wavelength ($k_z<1/2H$) are transiently stable against the Parker
instability, during a few shear times. On the contrary, slow MHD waves with
a short vertical wavelength are destabilized by differential rotation almost
as long as their polarization is in the plane of the disc, \ie while
$|K_x(t)|<k_z$. The Parker instability naturally occurs when the
polarization of the slow MHD wave is vertical, \ie when $|K_x(t)|>k_z$.

\par (v) A refined look at the transition between the Parker and shearing
instabilities shows that the intensity of the differential rotation is the
crucial parameter. If the rotation curve decreases more slowly than the flat
rotation curve ($-A/\Omega<1/2$), the growth rate will always vanish between
the two instabilities. If the rotation curve decreases faster than the flat
rotation curve ($-A/\Omega>1/2$), the two instabilities can occur
simultaneously: in discs with a strong differential rotation, the most
unstable perturbations may not be the ones with $k_z=0$, because of this
shearing instability.\\

Astrophysical and observational consequences of these results are beyond the
scope of this paper, which leaves many questions open: observing shearing
instabilities might prove even more difficult than looking for the signature
of Parker instabilities, because here one is dealing with small vertical
wavelengths, on which the gas disc and the field are hardly homogeneous.
This might actually be seen as an obstacle to the development of the
instability, or as a consequence of it (see the discussion in Zweibel \&
Kulsrud 1975, Parker 1975b).\\
Furthermore, since we are limited to linearized
analysis, we cannot answer the question of the ultimate fate of these waves.
If they reach very large amplitudes before the shear takes them to very
large $K_x$ they might more easily, through the non-linear effect of
self-gravity on the resulting clumps of gas, lead to Jeans collapse of these
clumps. On the other hand if they reach large $K_x$ their ultimate fate
would be more likely a cascade towards small-scale turbulence.\\
One should also remember, in such a discussion, that in galaxies the gas disc
is swept and shocked by the spiral arms on a time scale of the order of a
rotation time, \ie at best a few amplification times, so that the coherent
growth of a Parker or Shearing instability on a longer time scale is rather
dubious. The conclusions would obviously depend on assumptions on the initial
state, \ie the amplitude of fluctuations generated in the disc and ready to be
amplified by these instability mechanisms.\\

On the other hand self-gravity might play a quite different role, which we
consider as most promising for generating strong Parker-like perturbations
in the disc: here and in [FT] we have shown that differential rotation
results in a linear coupling of waves, \ie a wave on a given branch of the
dispersion relation evolves into a mixture of the three MHD waves (slow and
fast magnetosonic, and Alfven waves). If we had included self-gravity in
this discussion, the fast magnetosonic wave would have appeared as the
classical spiral density waves of galactic discs, modified by inclusion of
the magnetic field (as was done, in the case of a vertical field, by Tagger
\etal 1990). But one can conclude that a spiral wave does, by this mechanism
of linear coupling, generate Alfven and slow magnetosonic waves, \ie Parker
rather than shearing instabilities since this occurs at $k_z=0$. We know
that the spiral wave is strong (the density contrast is of the order of
100$\%$ in the gas), and the coupling is also strong since the rotation
curve has $-A/\Omega\sim .5$. This means that the resulting Parker
perturbation will also have a very large amplitude, even before it has
started to grow by Parker's mechanism. We consider this as the most
promising way of generating strong Parker perturbations in the disc, and
will discuss it in future work.

\begin{acknowledgements}
We thank Prof. R. Pudritz for his useful comments. T.F. wishes to
acknowledge fruitful discussions with Dr. H. Spruit and Dr. F. Meyer.
\end{acknowledgements}

\appendix
\section{Appendix: Laplace Transform of the solution with shear}
We define the dimensionless rotation parameter and epicyclic frequency as:
\begin{eqnarray}
\tilde\Omega^2&\equiv&{4H^2\Omega^2\over \VAc},\\
\tilde\kappa^2&\equiv&{\kappa^2H^2\over \VAc}=\tilde\Omega^2-\bk_A^2.
\end{eqnarray}
In order to reduce the degree of the differential system
(\ref{systdiff3}), we introduce the convenient function:
\begin{eqnarray}
\tx_w=-{1+2\alpha\over2\alpha}\left\{\bar K_x\tx_x\right.\nonumber\\
\left.+{\bk_y\tx_y\over1+
2\alpha}+\left( \bk_z+i{1+2\beta\over2+4\alpha}\right)\tx_z\right\}
\end{eqnarray}
The only time-dependent parameter, $K_x$, now multiplies the functions
$\tx_x$ and $\tx_w$ only. Thus we can write two equations without any
$K_x$-multiplication, and two other equations expressing $K_x\tx_x$ and
$K_x\tx_w$ in terms of $\tx_w,\tx_x,\tx_y,\tx_z$. These latter are
written:
\begin{eqnarray}
\bar K_x \tx_w &=& \left({H^2\over \VAc}{\p^2\over\p
t^2}+\bk_y^2-\bk_A^2\right)\tx_x -\tilde\Omega{H\over \VA} {\p\over\p
t}\tx_y\label{Kwx1},\\
\bar K_x\tx_x &=& {-2\alpha\over 1+2\alpha}\tx_w- {\bk_y\over
1+2\alpha}\tx_y -  \left(\bk_z+i{1+2\beta\over 2+4\alpha}\right)\tx_z.
\label{Kwx2}
\end{eqnarray}
Performing at this point the Laplace transform (\ref{Laplace}) changes the
$K_x$-multiplications into $\omega$-derivatives, and time-derivatives change
to simple $\omega$-multiplications, with some additional
$\omega$-polynomials whose coefficients depend linearly on the initial
conditions $\tx(k_x,t=0)$ and ${\p\over\p t}\tx(k_x,t=0)$.\\
In the following, we use the dimensionless variable $\tom\equiv
H\omega/\VA$.\\
The first couple of equations, independent of $K_x$, relate ($\bx_y,\bx_z$)
to ($\bx_w,\bx_x$) in a simple algebraic manner:
\begin{equation}
\Delta
\left|\begin{array}{l}
\bx_y\\
\bx_z
\end{array} \right.
 =
\left(\begin{array}{cc}
q_a & q_b\\
q_c & q_d
\end{array} \right)
\left|\begin{array}{l}
\bx_w\\
\bx_x
\end{array} \right.
+\e^{i{\tom\tilde\Omega\bk_x\over\bk_y\bk_A^2}}
\left|\begin{array}{l}
Q_y\\
Q_z
\end{array} \right..\label{qabcd}
\end{equation}
$Q_y$ and $Q_z$ are polynomials of $\tom$ whose coefficients depend linearly
on the initial conditions. The polynomials ($q_a,q_b,q_c,q_d$) are
independent of $\bk_x$, and constitute a matrix which is invertible as long
as $\Delta\not=0$ and $\tom\not=0$:
\begin{eqnarray}
q_a&=&-\bk_y\left\{\tom^2-\bk_y^2
+(1+\alpha+\beta)\left(i\bk_z+{\beta\over2\alpha}\right)\right\},\nonumber\\
q_b&=&i\tilde\Omega \tom\left\{(1+2\alpha)(\bk_y^2-\tom^2)+
(1+\alpha+\beta){\alpha-\beta\over 2\alpha}\right\},\nonumber\\
q_c&=&{i\over2}\{\bk_y^2+(1+2\beta)\tom^2\}+\bk_z\{\bk_y^2-
(1+2\alpha)\tom^2\},\nonumber\\
q_d&=&-\tilde\Omega \tom\bk_y(1+\alpha+\beta).\nonumber
\end{eqnarray}
The polynomial $\Delta$, already introduced in [FT], identifies with the
asymptotic dispersion relation at $k_x\to\infty$:
\begin{eqnarray}
\Delta\!&\equiv&\!(1+2\alpha)\tom^4
-2\left\{(1+\alpha)\bk_y^2+(1+\alpha+\beta)
{\alpha-\beta\over4\alpha}\right\} \tom^2\nonumber\\
& &\hspace{2cm}+\bk_y^2(\bk_y^2-\bk_{\rm P}^2).
\label{delta}
\end{eqnarray}
The maximum azimuthal wavenumber $k_{\rm P}$ allowing the Parker instability
at $k_x\to\infty$ is defined as:
\begin{equation}
\bk_{\rm P}^2\equiv (1+\alpha+\beta){\alpha+\beta\over2\alpha}>{1\over2}.
\label{defkp}
\end{equation}
Equations (\ref{Kwx1})-(\ref{Kwx2}) may be expressed as a differential
system of second order on the functions  ($\bx_w,\bx_x$), using
Eq.~(\ref{qabcd}).
\begin{equation}
{\bk_A^2\bk_y\over\tilde\Omega}\Delta{\p\over\p \tom}
\left|\begin{array}{l} \bx_w\\ \bx_x \end{array} \right.
 =
\left(\begin{array}{cc}
p_a & p_b\\
p_c & p_d
\end{array} \right)
\left|\begin{array}{l}
\bx_w\\
\bx_x
\end{array} \right.
+\e^{i{\tom\tilde\Omega\bk_x\over\bk_y\bk_A^2}}
\left|\begin{array}{l}
P_w\\
P_x
\end{array} \right.,
\label{diffabcd}
\end{equation}
where ($p_a,p_b,p_c,p_d$) are simple polynomials of $\tom$ whose
coefficients are independent of $\bk_x$.\\
$P_w$ and $P_x$ are two polynomials of $\tom$ of respective degree 5
and 3, whose coefficients depend linearly on the six initial
conditions
\begin{equation}
\left(\tx_x,\tx_y,\tx_z,{\p\over\p t}\tx_x,{\p\over\p t}\tx_y,
{\p\over\p t}\tx_z\right)(k_x,t=0),\nonumber\end{equation}
independently from the $\bk_x$ and $\bk_A$ parameters.\\
The coefficients of the matrix in Eq.~(\ref{diffabcd}) are given
by:
\begin{eqnarray}
p_a&=&\tilde\Omega\bk_y \tom\left\{\tom^2-\bk_y^2
+(1+\alpha+\beta) \left({\beta\over2\alpha}+i\bk_z\right)\right\}\nonumber\\
p_b&=&-i\{(\tom^2-\bk_y^2 -\tilde\kappa^2)\Delta(\tom^2)
-\tilde\Omega^2k_y^2(\tom^2-\bk_y^2+\bk_{\rm P}^2)\}\nonumber\\
p_c&=&-iQ_2(\tom^2)-i\{\bk_y^2-(1+2\alpha)\tom^2\}\bk_z^2\nonumber\\
p_d&=&-\tilde\Omega\bk_y \tom\left\{\tom^2-\bk_y^2
+(1+\alpha+\beta) \left({\beta\over2\alpha}-i\bk_z\right)\right\}\nonumber
\end{eqnarray}
The determinant of (\ref{diffabcd}) is $p_ap_d-p_bp_c=\Delta (Q_3-\bk_z^2
Q_A)$.\\ Defining $\bk_{\rm Q}^2\equiv\bk_{\rm P}^2-1/4$ leads us to write
the polynomials $Q_2,Q_3$ and $Q_A$ as:
\begin{eqnarray}
Q_2&\equiv
&2\alpha\tom^4-(1+2\alpha)\left(\bk_y^2+{1\over4}\right)\tom^2+
\bk_y^2(\bk_y^2-\bk_{\rm Q}^2)\label{eq_2}\\
Q_3&\equiv&(\tom^2-\bk_y^2-\tilde\kappa^2)
Q_2(\tom^2)-k_y^2\tilde\Omega^2(\tom^2+\bk_{\rm
Q}^2-\bk_y^2)\label{dispshear}\\
Q_A&\equiv&(1+2\alpha)\tom^4-\tom^2\{2(1+\alpha)\bk_y^2+
(1+2\alpha)\tilde\kappa^2\}\nonumber\\
 & &\hspace{2cm}+\bk_y^2(\bk_y^2-\bk_{\rm A}^2).
\end{eqnarray}
The differential system (\ref{diffabcd}) can easily be transformed, by
substitution, into the second order differential equation:
\begin{equation}
\left\{{\p^2\over\p\tom^2}\!+\!{\p\log f(\tom)\over\p\tom}{\p\over\p\tom}\!
+\!g(\tom) \right\}\bx(k_x,\tom)
=h(k_x,\tom),\label{Frobenius}
\end{equation}
where $h$ depends linearly on the initial conditions, and ($f,g,h$) are
singular where $\Delta(\tom^2)=0$.\\
We derive for $\bx_x$:
\begin{eqnarray}
{\p\log f_x\over\p\tom}&=&{\p\log \Delta/p_c\over\p\tom}
-{\tilde\Omega\over\bk_A^2\bk_y}{p_a+p_d\over\Delta},\label{fx}\\
g_x(\tom^2)&=&{\tilde\Omega^2\over\bk_A^4\bk_y^2}\left\{{Q_3-\bk_z^2Q_A
\over\Delta}
-{\bk_A^2\bk_y\over\tilde\Omega}{p_c\over\Delta}{\p\over\p\tom}{p_d\over
p_c}\right\}.
\label{gx}
\end{eqnarray}

\section{Appendix: WKB approximation}
We concentrate here on the function $\bar\xi_x$, without specifying the
subscript $x$ for the sake of clarity. The homogeneous differential equation
associated to Eq.~(\ref{Frobenius}) may be written in its canonical form:
\begin{equation}
\left\{{\p^2\over\p\tom^2}+W(\tom^2)\right\}(f^{1/2}\bx_H)=0,
\label{canon}
\end{equation}
where $W(\tom^2)$ is a fractional function of $\tom^2$ defined as:
\begin{equation}
W(\tom^2)\equiv g(\tom^2)-{1\over2}{\p^2\log
f(\tom^2)\over\p\tom^2}- {1\over4}\left({\p\log
f(\tom^2)\over\p\tom}\right)^2.\label{defW}
\end{equation}
We have shown in [FT] that if the  WKB criterion is satisfied,
\begin{equation}
\left|{1\over4}{\p^2\log W\over\p\tom^2}-\left({1\over4}{\p\log
W\over\p\tom} \right)^2\right|\ll g,\label{criterWKB}
\end{equation}
the solutions of the differential equation (\ref{canon}) can be approximated
by a linear combination of the two independent functions:
\begin{equation}
\bx_\pm(\tom)\equiv {1\over (f^2W)^{1\over4}}\exp\left(\pm
i\int^{\tom} W^{1/2}(\tom'^2)\d\tom'\right).
\end{equation}
According to the definition (\ref{fx}) of the function $f$, we can write:
\begin{equation}
\bx_\pm(\tom)\equiv {p_c^{1/2}\over \Delta^{1\over2}W^{1\over4}}\exp
i \int^{\tom} \left({\tilde\Omega\over\bk_y\bk_A^2}
{p_a+p_d\over2i\Delta}\pm W^{1/2}\right).
\end{equation}
Let us define the phase $\Psi_\pm(\tom,t)$ in a dimensionless form as:
\begin{equation}
\Psi_\pm\equiv -{\tilde\Omega\over\bk_A^2\bk_y}\bar K_x(t)\tom+
\int^{\tom} \left({\tilde\Omega\over\bk_y\bk_A^2} {p_a+p_d\over2i\Delta}\pm
W^{1/2}\right).
\end{equation}
The solution of the differential equation (\ref{Frobenius}) is then written
as:
\begin{eqnarray}
\tx(k_x,t)={1\over2\pi}\int_{ip-\infty}^{ip+\infty}
\left(\mu_+\e^{i\Psi_+}- \mu_-\e^{i\Psi_-}\right)\d \omega,
\end{eqnarray}
where the two functions $\mu_\pm(\tom)$ vary slowly if the WKB criterion is
satisfied. The dispersion relation is therefore:
\begin{equation}
W(\tom^2)={\tilde\Omega^2\over\bk_A^4\bk_y^2}\left(\bar K_x(t)
-{p_a+p_d\over2i\Delta}\right)^2.
\end{equation}
The function $W(\tom^2)$ defined in (\ref{defW}) can be written:
\begin{eqnarray}
W=-{\tilde\Omega^2\over\bk_y^2\bk_A^4\Delta^2}\left\{p_cp_b+
\left(p_a-p_d\over2\right)^2 \right\} -{1\over2}{\p^2\over\p\tom^2}
\log{\Delta\over
p_c}\nonumber\\
+{\tilde\Omega p_c\over2\bk_y\bk_A^2\Delta}{\p\over\p\tom}{p_a-p_d\over p_c}
-{1\over4}\left({\p\over\p\tom}\log{\Delta\over p_c}\right)^2.
\end{eqnarray}
The general dispersion relation becomes:
\begin{eqnarray}
(p_a\!-\! i\bar K_x\Delta)(p_d\!-\! i\bar K_x\Delta)-p_cp_b
+{\bk_y\bk_A^2\Delta p_c\over2\tilde\Omega}{\p\over\p\tom}{p_a-p_d\over
p_c}\nonumber\\ -{\bk_y^2\bk_A^4\Delta^2
\over4\tilde\Omega^2}\left\{2{\p^2\over\p\tom^2}\log{\Delta\over
p_c}+\left({\p\over\p\tom}\log{\Delta\over p_c}\right)^2\right\}=0.
\end{eqnarray}
We recognize the first two terms as the determinant of the system
(\ref{diffabcd}). Some additional terms must be taken into account,
especially in the limit of both vanishing shear and magnetic field
(Sect.~5).\\ It can be divided by $\Delta$ to obtain:
\begin{eqnarray}
Q_3-Q_A\bk_z^2-\Delta\bk_x^2+2\tilde\Omega\tom(1+\alpha+\beta)
\bk_y\bk_x\bk_z\nonumber\\
+\bk_y^2\bk_A^2\left\{3\tom^2-\bk_y^2+(1+\alpha+\beta){\beta\over2\alpha}
\right\}\nonumber\\
-2\bk_y^2\bk_A^2\tom^2\left\{\tom^2-\bk_y^2+(1+\alpha+\beta)
{\beta\over2\alpha}\right\}{\p\log p_c \over\p\tom^2}\nonumber\\
-{\bk_y^2\bk_A^4\Delta\over4\tilde\Omega^2}\left\{2{\p^2\over\p\tom^2}
\log{\Delta\over p_c}+\left({\p\over\p\tom}\log{\Delta\over
p_c}\right)^2\right\}=0.
\label{dispers}
\end{eqnarray}
Both the WKB criterion (\ref{criterWKB}) and the dispersion relation
(\ref{dispers}) become much simpler in the two cases we investigate in
Sect.~(4-6):\\
\par (i) If the shear vanishes totally, we will obtain the usual
dispersion relation:
\begin{equation}
Q_3-Q_A\bk_z^2-\Delta\bk_x^2+2\tilde\Omega\tom(1+\alpha+\beta)
\bk_y\bk_x\bk_z=0. \label{classicdisp}
\end{equation}
\par (ii) If, as in Sect.~4 and Sect.~6, $k_z\to\infty$, we simply obtain:
\begin{eqnarray}
g(\tom^2)&\sim& {\tilde\Omega^2\over\bk_y^2\bk_A^4\Delta}Q_A(\tom^2)\bk_z^2,\\
W(\tom^2)&\sim& -{\tilde\Omega^2\over\bk_y^2\bk_A^4\Delta^2}
(-ip_b)(\bk_y^2-(1+2\alpha)\tom^2)\bk_z^2.
\label{dispgkz}
\end{eqnarray}
Thus the criterion (\ref{criterWKB}) is satisfied in the limit of large
vertical wavenumbers.\\
In Sect.~4, the radial wavenumber is bounded ($|K_x(t)|\ll |k_z|$), and the
dispersion relation (\ref{dispers}) becomes $Q_A(\tom^2)=0$.\\
In Sect.~6, $|K_x(t)|\sim |k_z|\gg 1/H$, and the dispersion relation
(\ref{dispers}) becomes:
\begin{equation}
Q_A+\left({\bar
K_x\over\bk_z}\right)^2\!\!\Delta-2\tilde\Omega\tom(1+\alpha+\beta)\bk_y
\left({\bar K_x\over\bk_z}\right)={\cal O}\left({1\over\bk_z^2}\right)
\end{equation}
\par (iii) If, as in Sect.~5, $\alpha\to 0$ and $\alpha\tilde\Omega^2\sim
cte$, then:
\begin{equation}
W(\tom^2)\sim g(\tom^2)\sim O(\tilde\Omega^4),
\end{equation}
and the WKB criterion is again satisfied. The simple dispersion relation
(\ref{frequency}) describing the slow MHD frequencies for finite values of
$k_z,K_x$ comes from the leading terms, when $\tom\tilde\Omega\sim 1$, of:
\begin{eqnarray}
Q_3-Q_A\bk_z^2-\Delta\bk_x^2+2\tilde\Omega\tom(1+\alpha+\beta)
\bk_y\bk_x\bk_z\nonumber\\
+\bk_y^2\bk_A^2\left(3\tom^2-\bk_y^2+(1+\alpha+\beta){\beta\over2\alpha}
\right)=0.
\end{eqnarray}

\section{Appendix: polarization of a slow MHD wave}
Here we recall the $(k_x,k_z)$-dependence of the polarization of a slow MHD
wave in absence of shear. Strictly speaking, the parameter $\bk_A$ ought to
be zero in what follows. \\
If $\tom$ is determined by the dispersion relation (\ref{classicdisp}), the
equations (\ref{diffabcd}) and (\ref{qabcd}) will imply:
\begin{equation}
{\xi_z\over\xi_x}={\bk_zQ_A-\tilde\Omega\tom(1+\alpha+\beta)\bk_y\bk_x+iR_1
\over \bk_x\Delta-\tilde\Omega\tom(1+\alpha+\beta)\bk_y\bk_z+iR_2},
\label{polarization}
\end{equation}
where $R_1,R_2$ are independent of $(\bk_x,\bk_z)$:
\begin{eqnarray}
R_1(\tom)&\equiv &\left({1\over2}+\beta\right)(\tilde\kappa^2-\tom^2)\tom^2
+\beta\bk_y^2\tom^2+{\bk_y^2\over2}(\bk_y^2-\bk_A^2),\\
R_2(\tom)&\equiv &\tilde\Omega\tom\bk_y\left[\tom^2-\bk_y^2+
(1+\alpha+\beta){\beta\over2\alpha}\right].
\end{eqnarray}
Let us study the polarization of the Alfven mode: we have seen in Sect.~4
that its eigenfrequency is a root of $Q_A(\tom^2)=0$ when the vertical
wavenumber $k_z\to \infty$, and we know from [FT] that it is a root of
$\Delta(\tom^2)=0$ when $k_x\to\infty$. Unlike the magnetosonic
eigenfrequency (fast MHD wave), it remains finite in both cases. Thus this
calculation is valid for both the Alfven and the slow MHD mode.\\
In the limit of $\bk_x\to\infty$, the dispersion relation
(\ref{classicdisp}) shows that:
\begin{equation}
\bk_x\Delta\sim 2\tilde\Omega\tod(1+\alpha+\beta)\bk_y\bk_z.
\end{equation}
The polarization (\ref{polarization}) is then:
\begin{equation}
{\xi_z\over\xi_x}\sim{-\tilde\Omega\tod(1+\alpha+\beta)\bk_y
\over \tilde\Omega\tod(1+\alpha+\beta)\bk_y\bk_z+iR_2(\tod)}\bk_x\to\infty,
\label{grandkx}
\end{equation}
corresponding to a vertical mode.\\
In the limit of $\bk_z\to\infty$, the dispersion relation (\ref{classicdisp})
implies:
\begin{equation}
\bk_zQ_A\sim 2\tilde\Omega\tom_s(1+\alpha+\beta)\bk_y\bk_x.
\end{equation}
The polarization (\ref{polarization}) becomes:
\begin{equation}
{\xi_z\over\xi_x}\sim{\tilde\Omega\tom_s(1+\alpha+\beta)\bk_y\bk_x+iR_1(\tom_s)
\over -\tilde\Omega\tom_s(1+\alpha+\beta)\bk_y}{1\over\bk_z}\to 0,
\label{grandkz}
\end{equation}
and corresponds to a horizontal mode.\\
If both $k_x,k_z\gg k_y$, the polarization (\ref{polarization}) is simply:
\begin{equation}
{\xi_z\over\xi_x}\sim -{\bk_x\over \bk_z}.
\end{equation}
Thus the plane of polarization of the slow MHD (and Alfven) mode is naturally
vertical or horizontal, depending on the ratio $k_x/k_z$.

These properties are specific of the slow magnetosonic mode {\it in a
stratified rotating disc}. In a uniform plasma ($H\to \infty$ and $\Omega\to
0$), the polarization of the slow magnetosonic mode is indeed completely
different, and is exactly:
\begin{equation}
{\xi_z\over\xi_x}={k_z\over k_x}
\end{equation}
It can easily be shown that:
\par(i) the property (\ref{grandkx}) is an effect of stratification, and is
valid only if $k_x\gg Hk_y^2$,
\par (ii) the property (\ref{grandkz}) is an effect of rotation, and is valid
only if $k_z\gg k_y^2a_s^2/\Omega$.

\end{document}